# Theoretical opportunities for rural innovation and entrepreneurship research

**ABSTRACT**

*Even though rural entrepreneurship and innovation has been studied for decades, the advent of social media, mobile, analytics, cloud computing and internet of things – also referred as digital technologies – (Nambisan 2013; Yoo et al. 2012) has provided new opportunities and challenges for this vast discipline. As a result, we see new business models, new processes, products and services offered using new digital technologies. Such changes challenge the orthodox view of IT entrepreneurship and innovation, opening new avenues for researches and challenges the existing theoretical understanding. This book chapter is an attempt to understand the existing literature on rural innovation and entrepreneurship in information systems discipline and identify opportunities for rural entrepreneurship and innovation in the digital era.*

## INTRODUCTION

The advent of digital technologies has provided unprecedented opportunities for all organizations across the globe, including those that could not afford to invest extensively in technologies that lead innovations (Nambisan 2013; Nylén and Holmström 2015; Sedera et al. 2016; Yoo et al. 2012). The innate characteristics of digital technologies such as cost effectiveness, ease of acquisition, ease of learning, ease of deployment and ease of management have made it easy for any organization to innovate using technologies (Lokuge and Sedera 2018; Nylén and Holmström 2015; Sedera and Lokuge 2017). Similarly, the advent of such technologies has created myriad opportunities for innovation and entrepreneurial activities in the rural areas (Cui et al. 2019; Li et al. 2019; Lokuge et al. 2016). While researchers such as Salemink et al. (2017) highlight the disparities of rural versus urban with regard to digital infrastructure, less attention has been paid to the opportunities available for minimizing the fundamental disadvantages in rural communities using digital technologies. As per Newbery et al. (2017), studying rural innovation and entrepreneurship will not only be valuable to academia, but it will add value to the practitioners such as rural communities, governments, small business holders and not-for-profit organizations in alleviating poverty, empowering women, improving economic conditions and minimizing the power disparities. Considering the advantages of contributing to such an interdisciplinary research area like rural innovation and entrepreneurship, the objective of this book chapter is to investigate the theoretical frameworks utilized in rural innovation and entrepreneurship studies in information systems (IS) context and thereby identify possible future research areas for IS researchers.

While researchers have studied the use of digital technologies in the developing countries context (e.g., Thaker et al. 2017; Venkatesh et al. 2016a; Venkatesh et al. 2016b), the context rurality is different to the context of developing countries. Therefore, it is important to define rurality in order to understand rural innovation and rural entrepreneurship. Most of the studies attempt to define rurality through social, economic and ecological components. For example, according to Dabson (2005) rurality can be defined through the differences in economy, values, environment and atmosphere. Further, studies such as Cloke (2006) conceptualize rurality using areas where most of the agricultural activities happen. While dissenting to the conceptualization of Cloke (2006), Newbery et al. (2017) proposes conceptualizing rurality through contextual factors such as peripherality from the center, uneven levels of development in terms of poverty, inequality and access to resources. Considering the prior conceptualizations, in this book chapter, rurality will be theorized by following the view of Stathopoulou et al. (2004), whereby rurality is determined through the three components such as geographic, social and economic.

According to Newbery et al. (2017), rural innovation and entrepreneurship plays a key role in promoting economic advances, providing more job opportunities and enhancing the livelihood of communities in the rural areas. Considering its importance, this book chapter is an attempt to contribute to the academia by providing an analysis on the extant theoretical frameworks on rural innovation and entrepreneurship. Through an archival analysis of major IS journals, this book chapter attempt to provide future research areas for IS researchers.

The remainder of this book chapter proceeds as follows. First, an overview to rural innovation and entrepreneurship is provided. Then, details of the literature review process are provided under the research methodology section. The analysis section provides the results of the literature review and a summary of theoretical foundations utilized in extant literature on rural innovation and entrepreneurship. Then, the theoretical framework for future research is provided. Finally, this article concludes by paving a path with an outlook on our next scholastic direction.

## RURAL INNOVATION AND ENTREPRENEURSHIP

This section provides an overview to rural innovation and rural entrepreneurship literature. In order to provide an overarching view of rural innovation and entrepreneurship, not only IS research, but also management, innovation and entrepreneurship disciplines have been studied in this section.

### Rural Innovation

Prior research alludes innovation to be a complex subject due to its mystical nature of its creation and adoption within an organization (Van de Ven 1986). Considering its importance and relevance, innovation has received attention of both academics and practitioners (Ahuja et al. 2016; Kohli and Melville 2019; Lokuge et al. 2019; Tan et al. 2016a). Innovation is considered as lifeblood or the survival mechanism for organizations in the contemporary competitive world (Leifer et al. 2000; Lewis et al. 2002; Utterback 1994). Though innovation is an extensively studied area, researchers posit various interpretations regarding innovation (Baregheh et al. 2009; Sears and Baba 2011). For example, there is a continuous debate among scholars in identifying innovation as a totally new idea or an imitation (Ruttan 1959). For most organizations, the term 'innovation' does not resonate with the 'new-to-the-world' concept. Researchers such as Lai et al. (2009) and Lyytinen and Rose (2003) argue that innovation need not be a totally new concept to the world and could even be considered as an imitation of something already used elsewhere, but new to the unit of adoption. For majority of organizations, innovation is not always an invention (Utterback 1971). As such, in this book chapter the latter conceptualization of innovation as an 'imitation of something already used elsewhere, but new to the unit of adoption' is followed. Following this, rural innovation is defined as the creation and implementation of new ideas or solutions, new to the unit of adoption, that deal with rural context (Sonne 2010).

When rural innovation is considered as an innovation, such conceptualization provides an opportunity to study it through widely classified typologies of innovation such as product innovation, process innovation, technical innovation and administrative innovation (Benner and Tushman 2003; Damanpour 1987; Damanpour 1991). Since the rurality in rural innovation encompasses its impacts on the social systems, as Pato (2015) suggested there are further extensions that are available for researchers when studying rural innovation. Further, with the advent of digital technologies, the accessibility to these technologies have provided unprecedented opportunities for small business owners in rural regions to innovate like resourceful counterparts. As a result, there is an opportunity for IS researchers to provide their expertise in digital innovation, innovation readiness, digital technology use, technology-led innovation and innovation adoption in the rural context.

**Rural Entrepreneurship**

Nambisan (2017, p. 1029) states that "digital technologies have transformed the nature of uncertainty inherent in entrepreneurial processes and outcomes as well as the ways of dealing with such uncertainty." Such discourses highlight the very nature of digital technologies and how they are changing the way entrepreneurs initiate their new business ideas. The relatively inexpensive, on-demand, functionally oriented, flexible digital technologies provide entrepreneurs with more opportunities to innovate (Lokuge and Sedera 2016; Lokuge et al. 2018; Nylén and Holmström 2015; Zittrain 2006). Such characteristics of digital technologies curtail the traditional barriers for entrepreneurship (Nambisan 2017). As such, rural communities are provided with an opportunity to become entrepreneurs (Tan et al. 2016b). Prior IS researchers have investigated common issues in IT use in developing countries (Leong et al. 2017; Leong et al. 2016b). In addition, such researchers have focused their attention on entrepreneurship in the growth stage of the entrepreneurial lifecycle (Tumbas et al. 2017; Wright and Stigliani 2013). Further, with the advent of new technologies such as mobile and social media, it is interesting to investigate how entrepreneurs in rural areas have utilized these platforms to identify new opportunities (Fischer and Reuber 2014). As such, the use of digital technologies in rural entrepreneurship context is an important yet, under-studied topic in the IS discipline.

The rural areas are inherently challenged with heterogeneous economic structures, government instability, weak and unstable economic institutions and limited skills and knowledge regarding business and technologies (Ernst 2002; Gupta 2013; Sedera et al. 2014; Tate et al. 2013). However, entrepreneurial activities play an important role in sustaining the socio-economic growth in rural areas (Pato 2015; Xiao et al. 2013). While rural entrepreneurship has been a much discussed topic of interest among researchers (Gaddefors and Anderson 2019; Newbery et al. 2017; Pato and Teixeira 2016; Wortman Jr 1990; Zahra 2007), most of the research on rural entrepreneurship has been conducted by relating rural context to entrepreneurship (Gaddefors and Anderson 2019). However, Stathopoulou et al. (2004) suggest that the geographic, social and economic contextual factors will provide different theoretical implication from this unique entrepreneurial genre. As such, it is interesting to analyze the nature of studies conducted in IS regarding rural entrepreneurship. Given the importance of rural entrepreneurship, it is important to understand theoretical foundations applied in the extant literature. By doing so, this book chapter attempt to provide new areas of research for rural entrepreneurship.

**RESEARCH METHODOLOGY**

This section of the book chapter details the methodology used in collating, synthesizing and analyzing the rural innovation and entrepreneurship literature. It outlines the search strategy and the journals reviewed for the analysis. The review of literature on rural innovation and entrepreneurship includes studies published between 1st of January 2009 to 31st of December 2019. Table 1 lists all journal outlets in IS used for the literature review. The selection of journals was inspired by the basket of eight journals list. Appendix A includes references of all the papers reviewed in the archival analysis from 2009 – 2019. The selection of the papers preferred inclusion, over exclusivity. In here, while searching for papers, even though the objective is to identify the application of digital technologies in the rural context, no specification was made for identifying papers on digital technologies in the rural context. That allowed us to canvass substantial body of developing literature. Each article identified through the search was read in its entirety and the relevance was determined. The most relevant papers were identified and then the theoretical frameworks of the papers were identified. The following keywords were used in the search "rural, developing nation*, developing countr*, developing econom*." The papers that were published from 2009 – 2019 in the last decade was considered for the analysis. However, for 2009, no paper related to rural innovation or entrepreneurship was found. At the end, the study included 22 papers that were considered relevant for the analysis.

*Table 1. Journals taken for the analysis*

| Acronym | Journal Name |
|---|---|
| MISQ | Management Information Systems Quarterly |
| ISR | Information Systems Research |
| JMIS | Journal of Management Information Systems |
| JAIS | Journal of Association for Information Systems |
| JIT | Journal of Information Technology |
| ISJ | Information Systems Journal |
| JSIS | Journal of Strategic Information Systems |
| EJIS | European Journal of Information Systems |

## ANALYSIS

This section provides the results of the analysis. Table 2 below depicts the number of papers used for the archival analysis. Out of the 22 papers, most of the papers were on e-commerce adoption in the rural context. Except for 4 papers, all 18 papers followed a case study approach.

*Table 2. Breakdown of the papers taken for the analysis*

| Journal | 2019 | 2018 | 2017 | 2016 | 2015 | 2014 | 2013 | 2012 | 2011 | 2010 | 2009 |
|---|---|---|---|---|---|---|---|---|---|---|---|
| MISQ | 0 | 0 | 0 | 1 | 1 | 0 | 1 | 0 | 0 | 0 | 0 |
| ISR | 0 | 0 | 0 | 1 | 0 | 0 | 1 | 0 | 0 | 0 | 0 |
| JMIS | 0 | 0 | 0 | 0 | 0 | 0 | 0 | 0 | 0 | 0 | 0 |
| JAIS | 1 | 1 | 1 | 0 | 0 | 1 | 0 | 0 | 0 | 0 | 0 |
| JIT | 0 | 0 | 0 | 0 | 0 | 0 | 2 | 0 | 0 | 0 | 0 |
| ISJ | 2 | 1 | 0 | 0 | 1 | 1 | 0 | 0 | 1 | 0 | 0 |
| JSIS | 1 | 0 | 1 | 0 | 1 | 0 | 0 | 0 | 0 | 0 | 0 |
| EJIS | 0 | 0 | 0 | 0 | 0 | 0 | 0 | 0 | 1 | 1 | 0 |

Additionally, an analysis based on the country where data is sourced from was conducted. For example, India, where 69% of the population live in rural areas had only 6 studies related to rural innovation and entrepreneurship in the sample. China was the second most frequently referred country, followed by Ghana, Kenya, Georgia and Australia. Overall, the results of the analysis highlights that there is a clear lack of focus on rural context and lot of opportunities for researchers to conduct new research.

In addition, the theories that have been applied was identified and Table 3 provides a summary of all the theories used in the review. Based on the analysis, 16 theories/frameworks were identified. The Arab policy and IT (APIT) model (Hill et al. 1998) was not considered for the analysis as it is not a theory. Activity theory (Leont'ev 1978) was the most commonly used theory in the sample. In addition to this, Resource orchestration perspective (Sirmon et al. 2011) and Frame theory (Cornelissen and Werner 2014) were commonly applied by researchers. The analysis of the sample highlighted that, most studies employed organizational-level theories and that only a handful of studies considered individual or group-level theories. The predominant focus on employing organizational level theories is understandable given that the overall focus of rural innovation initiatives tends to be at the organizational level. However, by focusing on individual-level adoption, proliferation, management and the benefits of rural entrepreneurship initiatives, practitioners could get better insights.

*Table 3. Identified theories and frameworks*

| Theory | Explanation/paper brief | Reference |
|---|---|---|
| Activity theory (Leont'ev 1978) | As per activity theory, human activity is considered as a subject (a person or a group) acting upon an object (the problem, situation, or focus of the activity), with the activity being mediated by means of | (Slavova and Karanasios 2018) |
| | material artifacts. The artifact is also referred to as tools. In this paper, by applying activity theory, the paper highlights how changes in the information and communication technologies among smallholder farmers in Ghana, led to a process of hybridization of information practices. The study also investigates the changes introduced as a result of using information and communication technologies. | |
| | This paper applies activity theory to examine how development actors within the Ghanaian agricultural sector endorse information technology in their day-to-day farming practices in the rural areas. | (Karanasios and Slavova 2019) |
| | The paper applies activity theory to understand the role of governments in rural China in managing rural e-commerce ecosystems and the impact of such ecosystems on poverty alleviation. | (Li et al. 2019) |
| Actor network theory (Callon 1999) | The theory posits that networks comprise of both social and technical parts and these elements are treated as inseparable by the theory. This paper utilizes actor network theory to explain ecommerce adoption in a rural context, which can be considered as an innovation. By utilizing actor network theory and unified theory of acceptance and use of technology, the paper proposes a global information technology adoption model. | (Datta 2011) |
| Effectuation theory (Sarasvathy 2001) | This theory describes how social entrepreneurs act in resource constrained environments. In a resource-constrained environment, the entrepreneurs follow an intuitive decision-making process, sometimes disregarding the rational choices on offer. This paper applies effectuation theory to understand the process of impact sourcing, investigate motivational triggers of impact sourcing entrepreneurship, the entrepreneurial actions underpinning different phases of venture creation and the positive institutional-level influences on impact sourcing of Indian companies. This theory has been widely used in entrepreneurship studies. | (Sandeep and Ravishankar 2015b) |
| Frame theory (Cornelissen and Werner 2014) | Frame theory posits that frames are socio-cognitive structures that individuals use to make sense of the world. Through these frames, policymakers understand the problems and identify possible solutions. This paper through the application of frame theory, investigates how incongruent frames evolve over time in IS innovation and what implications and changes the adoption of IS have on innovation processes in the rural healthcare innovation setting. | (Bernardi et al. 2017) |
| | By applying frame theory, this paper investigates how impact sourcing ventures frame their activities to marginalized communities. The paper provides an overview to the difficulties faced by impact sourcing ventures in operationalizing their strategic intent and how different and diverse framings are used by impact sourcing ventures to influence the local rural community. | (Sandeep and Ravishankar 2015a) |
| Innovation diffusion theory (Rogers 1995) | The innovation diffusion theory explains why, and at which rate new ideas and technologies spread. The theory posits that the innovators are eager to adopt new technologies; however, individuals who are risk-averse, cautious and skeptical are more likely to be laggards, in terms of adopting technology compared to innovators. Usually there is a general mistrust among governments and communities in rural areas. | (Venkatesh et al. 2014) |

| | As such, there is a reluctance among rural communities in adopting eGovernment initiatives, which can be considered as an innovation for the rural communities. | |
|---|---|---|
| Institutional theory (DiMaggio and Powell 1983) | Institutional factors have been applied to understand e-government policies and initiatives undertaken by the Government of Jamaica. The institutional theory provides the theoretical underpinning to understanding the socially constructed evolving logics that give meaning to organizational actions. This theory has been applied in understanding technological innovations. In this paper, the institutional theory is applied to understand a rural telehealth innovation and to understand the organizational transformations and path constitution patterns. | (Brown and Thompson 2011) |
| Market efficiency theory (Kendall and Dearden 2018) | The market efficiency theory has been applied to investigate the use of contemporary technologies such as mobile technologies in agri-business to get agriculture related information, market information and to identify market trends. The theory posits that the key problem of the farmer is the inability to gain a fair price for their products. According to the theory, the adoption of technologies for agriculture can enhance the participation of the farmers and thereby can get a greater income for their produce. | (Karanasios and Slavova 2019) |
| Path constitution theory (Van de Ven and Poole 1995) | This theory combines path dependence and path creation perspectives. Path dependence focuses on historically embedded, contingent processes that are beyond the control of the actors. Path creation focuses on mindful contributions from powerful actors. The paper applies path dependence and path creation perspectives to understand how technological innovation paths constitute in organizational contexts. | (Singh et al. 2015) |
| Postcolonial theory (Adam and Myers 2003) | The theory provides an understanding to the ways in which colonialism continues to affect the former colonies after political independence. The theory emphasizes the negative effects of power asymmetries on development projects in former colonies (developing countries). The asymmetric power relationship can be applied to the rural context as well. This theory can be applied to the rural context as this covers contextual factors such as culture, leadership, identity and power that defines the rurality. This can be applied to both rural innovation and entrepreneurship context. | (Chipidza and Leidner 2019) |
| Resource orchestration perspective (Sirmon et al. 2011) | Resource orchestration theory extends the resource based view (Barney 1991) by proposing that resources alone might not have rare, inimitability, valuable, non-substitutable characteristics that promote competitive advantage. However, orchestration of resources can provide such characteristic which may lead to gain competitive advantage. This paper applies orchestration perspective to investigate how social entrepreneurs in rural China orchestrate resources to achieve social innovation and it explains the processes for achieving social innovations in the rural context. | (Cui et al. 2017) |
| | This paper investigates community capability, a capability developed through a lead user's individual capability and the interactions between bundling product and e-commerce knowledge resources by rural community members and structuring institutional and legal resources | (Cui et al. 2019) |

| | by the local government. The authors further investigated the relationship between the individual capability of a lead user and the interactions between the resource orchestrations of different stakeholders and community capability using the resource orchestration perspective. | |
|---|---|---|
| Social identity theory (Tajfel 1981) | Social identity theory postulates that social category (e.g., nationality, political affiliation) with which an individual feels a sense of belonging provides a definition of that individual. However, an individual may feel belonging to multiple social categories. Through the application of this theory and social markedness theory, the paper suggests two mechanisms by which functional limitations affect a digitally disadvantaged person's adoption decision. They are: (i) technology perceptions (i.e., perceived usefulness, perceived ease of use, and perceived access barriers), and (ii) marked status awareness (i.e., stigma consciousness). | (Pethig and Kroenung 2019) |
| Social markedness theory (Brekhus 1998) | Social markedness theory holds the view that social actors actively perceive one side of a contrast while ignoring the other side as epistemologically unproblematic. In the paper, marked individuals are considered as those who are digitally advantaged. Through the application of this theory and social identity theory, the paper provides two mechanisms by which functional limitations affect a digitally disadvantaged person's adoption decision. They are: (i) technology perceptions and (ii) marked status awareness. | (Pethig and Kroenung 2019) |
| Social network theory (Wasserman and Faust 1994) | Social network theory provides a theoretical lens to understand how individuals' ties with others in a given network. This paper applied social network theory to understand the benefits of a digital divide intervention that is influenced by the power and the actors in the network. | (Venkatesh and Sykes 2013) |
| Theory of market separations (Bartels 1968) | The market separations theory describes barriers in the flow of goods, information, and money that obstruct the market exchanges between producers and consumers. The theory provides four types of disconnections between producers and consumers. They are: Spatial, Temporal, Informational, and Financial. By utilizing this theoretical perspective, the paper investigates how and why ICT-enabled innovations in products and processes deployed for market development at the bottom of the pyramid (individuals whose income is less than $2 per day), enable developmental outcomes through reduction of market separations. | (Tarafdar et al. 2013) |
| Unified theory of acceptance and use of technology (UTAUT) (Venkatesh et al. 2003) | The unified theory of acceptance and use of technology theory was developed through a review and union of the constructs of eight models such as theory of reasoned action, technology acceptance model, motivational model, theory of planned behavior, a combined theory of planned behavior/technology acceptance model, model of PC utilization, innovation diffusion theory, and social cognitive theory. The UTAUT attempts to explain user intentions to use an IS and subsequent usage behavior. According to UTAUT, four key constructs such as performance expectancy, effort expectancy, social influence, and facilitating conditions are considered as determinants of usage intention and behavior. | (Datta 2011) |

# FUTURE RESEARCH DIRECTIONS

Figure 1 provides the proposed framework for studying rural innovation and entrepreneurship using digital technologies. Figure 1 summarizes the future research areas for rural innovation and entrepreneurship literature.

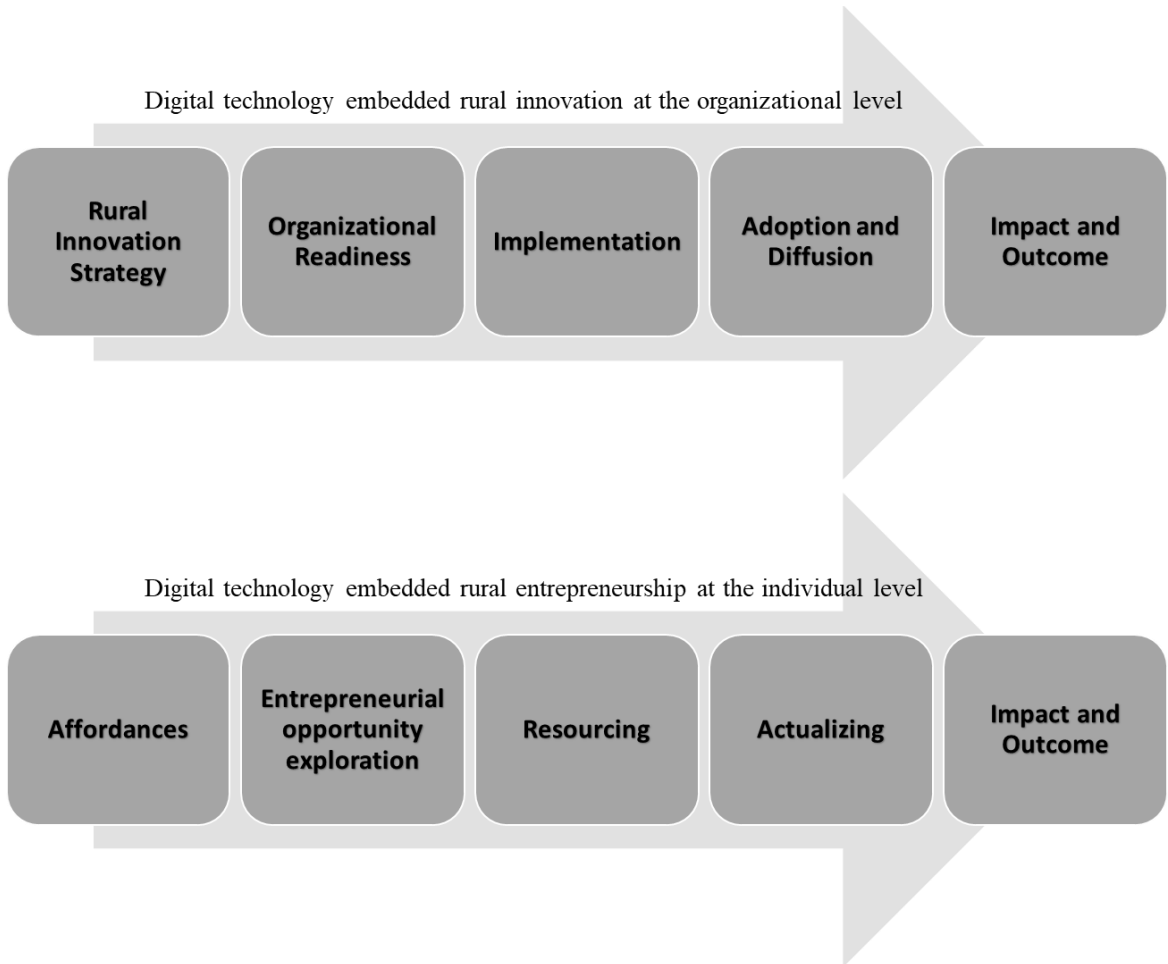

*Figure 1. Rural innovation and entrepreneurship future research areas*

When considering the theories utilized in rural innovation and entrepreneurship context in IS discipline, it is evident that IS researchers have myriad of opportunities in this area. For example, when considering theories such as the technology diffusion theory of Rogers (1995), even though this has been used in rural innovation context, this theory requires a rethink. As per Brown and Thompson (2011) an organic process subject to market push-pull mechanisms, is not sufficient for rural innovation context as in the rural context, not only the organizational level factors, but also many other individual and societal factors hinder the diffusion of rural innovation. Investigation of such factors will add value not only to the academia, but also to the practitioners. In addition, technology acceptance model (Davis 1989), theory of planned behavior (Ajzen 1991) and theory of reasoned action (Ajzen and Fishbein 1973) can be extended to rural innovation context, applying to the rurality context. The technology acceptance model (Davis 1989) provides theoretical explanation to better understand the innovation adoption behaviors of individuals (Salim et al. 2014). The individuals in a rural context differs from economical, geographical and social aspects. As such, IS researchers could identify new factors that affects the adoption behavior of the rural community. The UTAUT (Venkatesh et al. 2003) can also be extended by similar to technology acceptance model (Davis

1989). Similarly, the innovation diffusion in rural community can be affected by different contextual factors as such, Rogers's innovation diffusion theory too can be applied to generate new insights.

While the development of new theories such as power parity theory (Chipidza and Leidner 2019) is applicable to developing nations and ICT4D (information and communication for development), the asymmetry in power can be applied to the rural innovation and entrepreneurship context as well. For example, even in a developed nation, the rural areas have disparity in resources, infrastructure and skills. As such, IS researchers could apply their expertise in contributing to academia and practice.

For rural entrepreneurship context, the application of chauffeur theory (Markus 2001) can add value. As per chauffeur theory (Markus 2001), the leaders are not only information coordinators, but also ecosystem orchestrators and the source of motivation. Since in a rural context, the entrepreneur is faced with resource limitations, this can be applied to understand the rural entrepreneurs' behavior.

Affordances captures the positivity the stakeholders have towards entrepreneurship and how they perceive digital technology in the context of rural entrepreneurship. The entrepreneurial exploration, resourcing, actualizing, can be applied to assess rural entrepreneurship. With the advent of digital technologies, it is interesting to see how these technologies affect each of these important aspects of entrepreneurship in the rural context.

Future studies of rural innovation and entrepreneurship can consider the below areas as well:

- **Rural innovation modes:** there are three types of innovation modes discussed in literature. For example, incremental, quantum, and disruptive innovation. The incremental innovation refers to gradual improvements. The quantum innovation refers to a major improvement to an existing initiative that displaces current understanding (Maxwell 2009). The disruptive innovation as the name suggests is a major improvement to an existing initiative. Similarly, researchers can identify different rural innovation modes and contextual factors for such innovation modes.

- **Rural innovation and entrepreneurship contextual factors:** Most of the rural innovation and entrepreneurship research to date has focused on studying adoption factors. The IS researchers could identify new contextual factors considering the novelty of the rurality context by conducting case studies and surveys.

- **Rural innovation strategies:** The IS researchers can utilize their years of expertise in technology led strategies, resource-based view and strategic management literature in identifying new strategies for rural context. Further, the role of the technology in rural innovation and entrepreneurship, leadership and the management of strategies for successful innovation can be further studied.

- **Dimensions of rural innovation and entrepreneurship:** prior research on innovation generally distinguish innovation by the speed of the innovation process and its frequency to deliver innovations to the market (Benner and Tushman 2003). Innovation speed can be defined as the duration of the ideation and the commercialization of an innovation (Kessler and Chakrabarti 1996). Innovation frequency determines how often an organization introduces and delivers new products and services to the market (Pettigrew et al. 2001). Similar concepts can be applied to rural innovation and entrepreneurship context to add value to academia and practitioners.

## CONCLUSION

In reviewing the academic papers, it was evident that rural innovation and entrepreneurship have not yet received its due diligence. In almost all countries around the world, there are rural areas where there is lack of infrastructure, lack of resources and geographically distant from the urban areas. According to World Bank, in 2015 45% of the population resides in rural communities in the world. Even though, the impact of rural innovation and entrepreneurship is high, this concept is new to many countries and communities. With the advent of digital technologies rural communities similar to urban communities are provided with myriad opportunities for innovation and entrepreneurial activities (Cui et al. 2019; Li et al. 2019). The objective of this book chapter was to investigate the application of digital technologies in rural innovation and entrepreneurship in information systems (IS) context and thereby identify the future research areas for IS researchers. To identify the theoretical opportunities, this book chapter analyzed papers on rural innovation and entrepreneurship published in top IS journals from 2009-2019. Based on the analysis, 16 theories/frameworks were identified and only 15 theories were utilized for the analysis. While most of these theories are on organizational level, there are lot of opportunities for future research areas. For example, there is an opportunity for IS researchers to provide their expertise in digital innovation, innovation readiness, digital technology use, technology-led innovation and innovation adoption in the rural context. Also, they can utilize different methodologies to assess the rural context as well (Sedera et al. 2003; Sedera et al. 2001). Further, the researchers can investigate environmental factors, organizational factors, technological factors and individual factors that affect rural innovation and entrepreneurship. Through the review of literature, it was evident that in IS outlets, none of the papers investigated innovation using digital technologies. As such, it was clear that there is a huge opportunity for IS researchers to contribute to research on rural innovation and entrepreneurship using digital technologies. Further, this book chapter provided a framework that can be utilized for IS researchers for identifying future research areas in rural innovation and entrepreneurship.

**ADDITIONAL READING**

**KEY TERMS AND DEFINITIONS**

**Digital technologies:** New technologies such as social media, mobile technologies, analytics, cloud computing and internet-of-things are considered as digital technologies.

**Rural innovation:** the creation and implementation of new ideas or solutions, new to the unit of adoption, that deal with rural context.

**Rural entrepreneurship:** the discovery, evaluation, and exploitation of novel ideas in the rural context.

**Theory:** a system of ideas intended to explain a phenomenon.

**Framework:** a conceptual structure for explaining a phenomenon

**Entrepreneurship:** the discovery, evaluation, and exploitation of future goods and services

**Innovation:** implementation of an idea whether pertaining to a device, system, process, policy, program or service that is new to the organization.

**APPENDIX 1: Papers taken for the analysis**

| Year | Reference | Count |
|---|---|---|
| 2010 | (Okoli et al. 2010) | 1 |
| 2011 | (Brown and Thompson 2011; Datta 2011) | 2 |
| 2013 | (Dobson et al. 2013; Foster and Heeks 2013; Tarafdar et al. 2013; Venkatesh and Sykes 2013) | 4 |
| 2014 | (Sanner et al. 2014; Venkatesh et al. 2014) | 2 |
| 2015 | (Sandeep and Ravishankar 2015a; Sandeep and Ravishankar 2015b; Singh et al. 2015) | 3 |
| 2016 | (Leong et al. 2016a; Venkatesh et al. 2016a) | 2 |
| 2017 | (Bernardi et al. 2017; Cui et al. 2017) | 2 |
| 2018 | (Slavova and Karanasios 2018) | 1 |
| 2019 | (Chipidza and Leidner 2019; Cui et al. 2019; Karanasios and Slavova 2019; Li et al. 2019; Pethig and Kroenung 2019) | 5 |